\documentclass[pra,aps,10pt,twocolumn,amsmath,amssymb]{revtex4-1}

\usepackage[english]{babel}
\usepackage[utf8]{inputenc}
\usepackage{afterpage}
\usepackage{graphicx}
\usepackage{dcolumn}
\usepackage{bm}
\usepackage{color}
\usepackage{slashed}
\usepackage{latexsym}
\usepackage{appendix}
\usepackage{esint}
\usepackage{enumitem}

\newcommand{\dd}{\mathrm{d}}         
\newcommand{\ii}{\mathrm{i}}  
\newcommand{\ee}{\mathrm{e}}  

\DeclareUnicodeCharacter{2009}{ }

\begin{document}

\title{Laser-assisted radiative recombination beyond the dipole approximation}
\author{Deeksha Kanti$^1$}
\author{M. M. Majczak$^1$}
\author{J. Z. Kami\'nski$^1$}
\author{Liang-You Peng$^{2,3,4}$}
\email{liangyou.peng@pku.edu.cn}
\author{K. Krajewska$^1$}
\email{Katarzyna.Krajewska@fuw.edu.pl}

\affiliation{$^1$Institute of Theoretical Physics, Faculty of Physics, University of Warsaw, Pasteura 5, 02-093 Warsaw, Poland \\
$^2$State Key Laboratory for Mesoscopic Physics and Frontiers Science Center for Nano-optoelectronics, School of Physics, Peking University, 
Beijing 100871, China \\
$^3$Beijing Academy of Quantum Information Sciences, Beijing 100193, China\\
$^4$Collaborative Innovation Center of Extreme Optics, Shanxi University, Taiyuan 030006, China}
\date{\today}

\begin{abstract}
A comprehensive theoretical approach to describe the electron-ion radiative recombination in the presence of intense, short laser pulses,
which accounts for nondipole corrections is presented. It is based on the relativistic Coulomb-Volkov solution describing an electron in
a combined Coulomb potential and a laser field, which is systematically expanded
in powers of $1/c$. Thus, it allows us to trace the origin of nondipole effects observed in the spectrum of emitted radiation.
Hence, as we demonstrate for high-frequency pulses assisting the process, a significant extension of the cutoff and asymmetry in angular 
distributions of the emitted radiation can be attributed to the electron recoil off the laser pulse. In addition, we investigate a possibility
of enhancing the efficiency of the generated high-energy radiation by chirping the pulse.
\end{abstract}

\maketitle

\section{Introduction}
\label{intro}

There exist different regimes of laser-matter interactions. For instance, the focus of strong-field physics is to describe laser-matter interactions 
in the nonrelativistic regime, in which the strength of the laser field is comparable to that binding electrons in atoms. For 
near optical fields, this concerns the intensity range of roughly $10^{13}$-$10^{17}$~W/cm$^2$. Beyond that, the electron dynamics in the
laser field becomes fully relativistic, which is already the subject of high-field physics. While the latter treats the laser field as 
a propagating electromagnetic wave, with a full account for its temporal and spatial dependence, in strong-field physics it is traditionally
described as a homogeneous time-dependent electric field. This being the essence of the dipole approximation. In more physical terms, 
in the dipole approximation the momentum transfer from the laser field and, hence, the radiation pressure
experienced by the electron are neglected. Finally, it is crucial to realize that while the strong-field physics relies on the time-dependent
Schr\"odinger equation, the correct framework for relativistic investigations is set by either the Dirac or the Klein-Gordon
equation. Hence, going beyond the standard strong-field physics framework leads to our better understanding of
processes at the crossover of both regimes.

Note that nondipole effects in strong-field phenomena have recently become a hot topic of scientific inquiry. A vast majority of those studies concern
ionization. For instance, it has been realized that the dipole approximation breaks down in describing ionization in high-frequency laser 
fields~\cite{Bugacov,Aldana,Forre1,Forre2,Forre3,Forre4,Forre5,Lei1,Suster1,Chen,Wang,Solve1,Suster2}. A detailed analysis has shown that the dipole approximation also fails for
the low-frequency but high-intensity laser fields~\cite{Reiss1,Reiss2}. The latter has been observed experimentally in the near-optical 
regime~\cite{exp1,exp2,exp3,exp4,exp5}. Initiated by those experiments, various aspects of laser-induced ionization beyond the dipole approximation
have been studied, aiming at a correct description of the underlying electron dynamics in a laser 
field~\cite{Titi,Chelkowski1,Ivanov,KK1,Chelkowski2,Chelkowski3,He,Keil,Hao,Lein1,Keitel,Boning,Lein2,Madsen1,Madsen2,Madsen3,Dejan1,Forre6} (see, also recent reviews~\cite{Pengreview,Kellerreview}).

Closely related to these are studies of the laser-assisted radiative recombination (LARR), which is the inverse process to the laser-induced ionization. 
Most importantly, both processes are of major importance as they constitute different stages of the high-order harmonic generation~\cite{HHG1,HHG2,HHG3}. 
We note that the yield of recombination can be significantly enhanced by a laser field. Qualitatively, this can be explained by using the 
classical picture. Namely, during one field period electrons are accelerated and decelerated by the field. Since, during their deceleration, 
there are much more favorable conditions for recombination to take place, the recombination yield increases. Here, a crucial question arises: 
How the features of LARR will be modified in conditions where the dipole approximation is not valid any more?
It is particularly interesting as in nonrelativistic considerations so far, LARR has been analyzed exclusively within the dipole 
approximation~\cite{Kami1,Kuchiev1,Kami2,Milo1,Leone1,Cheng,Biv,Bonanno,Frolov2,Seipt,Kami4,Milo6,Milo7,Tutmic,Multi,Hu,Kami5,Kami6,Leone3,
Leone4,Milo5,Deeksha}. We are aware of only one work regarding LARR in the relativistic framework~\cite{ChairaMuller}, treating the laser field as a propagating electromagnetic wave. More specifically, it concerned the radiative recombination by a near optical high-intensity laser field that was modeled as an infinite monochromatic wave. Also, it was assumed that the strong-field approximation (SFA) can be successfully applied; thus, disregarding the Coulomb electron-ion interaction. The latter is the essence of the Born approximation, which works very well in the relativistic settings, when the electron energy is high enough compared to the atomic binding potential. Under those circumstances, a very broad energy spectrum of emitted radiation was observed, that was saturated however at extremely relativistic electron energies. A characteristic shift in the angular distributions of emitted radiation was also found.

In contrast to Ref.~\cite{ChairaMuller}, in the current paper we investigate the LARR in short laser pulses of high-frequency. This calls for a new formulation of the process, which goes 
beyond the dipole approximation. We provide it in Sec.~\ref{theory} following the method introduced in Ref.~\cite{Deeksha}. While 
in Ref.~\cite{Deeksha} it was designed for the electron-atom radiative attachment, this time we account for the electron-ion Coulomb interaction. 
For this purpose, starting with the relativistic Klein-Gordon equation, we derive the Coulomb-Volkov state which is then analyzed in the 
leading order of expansion in $1/c$ (Sec.~\ref{VC}). In such way, nondipole contributions of different origins are recognized, which is different 
from typical approaches in strong-field physics. Having that, the probability amplitude of LARR is calculated (Sec.~\ref{amplitude}) for 
a monochromatic plane wave representing the initial electron in Sec.~\ref{monochromatic} and for a coherent superposition of such waves 
in Sec.~\ref{coherent}. The latter avoids the field-free singularities in the energy distributions of the emitted radiation, defined in 
Sec.~\ref{energy_distr}. In Sec.~\ref{numerics}, we present various numerical illustrations of our theoretical approach. Most importantly, 
we are able to trace the observed nondipole effects (such as the extension of the LARR cutoff and the asymmetry of angular distributions
of the emitted radiation) back to their origin. As it follows from our results (Sec.~\ref{nondipole}), they are mostly caused by the electron 
recoil off the laser pulse, which is confirmed by classical investigations of the electron dynamics in an electromagnetic field (Appendix~\ref{classical}). The remaining nondipole corrections, including the retardation correction, seem not to have a qualitative impact on the 
resulting LARR spectrum. Finally, in Sec.~\ref{control}, we investigate a possibility of increasing the intensity of the LARR harmonics by appropriately 
modifying the properties of the laser pulse. Our work is concluded
in Sec.~\ref{conclusions}, whereas details of our calculations are specified in Appendices~\ref{app} and~\ref{spect}.

Throughout the paper, in analytical formulas we set $\hbar=1$ while keeping the remaining fundamental constants explicitly. However, our
numerical results are given in atomic units (a.u.) of the momentum $p_0=\alpha m_{\rm e} c$, energy $E_0=\alpha^2m_{\rm e} c^2$, length
$a_0=\hbar/p_0$, time $t_0=\hbar/E_0$, electric field strength $\mathcal{E}_0=\alpha^3m_{\rm e}^2 c^3/(|e|\hbar)$, and the laser field intensity 
$I_0=\epsilon_0 c\mathcal{E}_0^2\approx 7.02\times 10^{16}\,\mathrm{W/cm}^2$, 
where $m_{\rm e}$ and $e=-|e|$ are the electron rest mass and charge, $\alpha$ is the fine-structure constant, and $\epsilon_0=e^2/(4\pi\alpha\hbar c)$ is the vacuum 
permittivity. 

\section{Theoretical formulation}
\label{theory}

In this section, we present the theoretical formulation of the laser-assisted electron-ion radiative recombination that is followed by a formation 
of a hydrogen-like atom and emission of a photon. Our objective is to introduce a systematic approach which accounts for nondipole corrections
of the order $1/c$. For this purpose, starting with the Klein-Gordon equation, we shall derive the Coulomb-Volkov states that describe the electron 
in a laser field. This relativistic solution will then be expanded in powers of $1/c$, thus providing the leading nondipole contributions to 
the probability amplitude of LARR. 


Consider recombination of an electron with an asymptotic momentum ${\bm p}$ by a Coulomb potential $V({\bm r})$ in the presence of a laser field; 
thus, resulting in emission of a photon characterized by the wave vector ${\bm K}$, energy $\omega_{\bm K}$, and linear polarization vector 
${\bm \varepsilon}_{\bm K}$. The probability amplitude of the process takes the form,
\begin{equation}
{\cal A}({\bm p})=-\ii \int_{-\infty}^\infty \dd t\,\langle \psi_B(t);1_{{\bm K}}|\hat{H}_I(t)|\psi_{\bm p}^{(+)}(t);0_{{\bm K}}\rangle,
\label{a1}
\end{equation}
where $|\psi_{\bm p}^{(+)}(t);0_{{\bm K}}\rangle$ is the initial state of the system representing the electron in the scattering state 
$\psi_{\bm p}^{(+)}({\bm r},t)$ and no photons, whereas $|\psi_B(t);1_{{\bm K}}\rangle$ is the final state of the system describing asymptotically the electron 
in the bound state $\psi_B({\bm r},t)$ of energy $E_B$ and an emitted photon. Moreover, $\hat{H}_I(t)$ represents the interaction Hamiltonian that causes 
the transition.

Note that both wave functions $\psi_{\bm p}^{(+)}({\bm r},t)$ and $\psi_B({\bm r},t)$ are, in principle, the exact electronic
wave functions. However, as we will argue below, for the initial wave function $\psi_{\bm p}^{(+)}({\bm r},t)$ it is justified to use the 
Volkov-Coulomb wave with nondipole corrections. Also, as we are dealing with short laser pulses, we shall assume that the final electron state 
$\psi_B({\bm r},t)$ is field-free.

\subsection{Volkov-Coulomb wave function}
\label{VC}
In this section, we derive the Coulomb-Volkov solution of the Schr\"odinger equation with the leading nondipole corrections of the order $1/c$. 
Our starting point is the Klein-Gordon equation for an electron coupled to an electromagnetic field,
\begin{equation}
\left[\frac{1}{c^2}(\ii\partial_t-V)^2-(\hat{\bm p}-e{\bm A})^2-(m_{\rm e}c)^2\right]\psi_{\bm p}({\bm r},t)=0.
\label{a1star}
\end{equation}
Thus, we neglect the electron spin.
Here, $V=V({\bm r})$ is the Coulomb potential, whereas ${\bm A}={\bm A}(t-{\bm n}\cdot {\bm r}/c)$ represents a vector potential 
describing a laser field. For the transverse laser field, that propagates in the direction ${\bm n}$, we assume that ${\bm n}\cdot {\bm A}=0$. 
We look for solutions of this equation in the form,
\begin{equation}
\psi_{\bm p}({\bm r},t)=\ee^{{\ii F}({\bm r},t)}\phi_{\bm p}({\bm r},t),
\label{a2}
\end{equation}
with the electron asymptotic momentum ${\bm p}$ and the phase depending on time and position such that $F({\bm r},t)=F(t-{\bm n}\cdot {\bm r}/c)$. 
Substituting the above solution to Eq.~\eqref{a1star}, we obtain the following differential equation for unknown functions $F({\bm r},t)$ and 
$\phi_{\bm p}({\bm r},t)$,
\begin{align}
\biggl[\frac{1}{c^2}(\ii\partial_t&-V)^2-\hat{\bm p}^2-(m_{\rm e}c)^2-\frac{2}{c^2}F'(\ii\partial_t-V)\nonumber\\
&+2\left(e{\bm A}+\frac{1}{c}F'{\bm n}\right)\cdot \hat{\bm p}-e^2{\bm A}^2\biggr]\phi_{\bm p}({\bm r},t)=0.
\label{a3}
\end{align}
Here, $F'$ means the derivative of the phase $F$ with respect to the retarded time, $t_{\rm R}=t-{\bm n}\cdot {\bm r}/c$. In order to simplify Eq.~\eqref{a3}, we further assume that
\begin{equation}
2\left(e{\bm A}+\frac{1}{c}F'{\bm n}\right)\cdot {\bm p}-e^2{\bm A}^2-\frac{2E_{\bm p}}{c^2}F'=0,
\label{a4}
\end{equation}
where $E_{\bm p}$ is the electron initial energy. Note that this condition can be rewritten in a relativistically invariant form by introducing the
four-vector notation. Taking $p=(E_{\bm p}/c,{\bm p})$, $n=(1,{\bm n})$, and $A=(0,{\bm A})$, Eq.~\eqref{a4} becomes
\begin{equation}
F'=-c\frac{eA\cdot p}{n\cdot p}+c\frac{e^2A^2}{2n\cdot p},
\label{a5}
\end{equation}
or, after integrating it,
\begin{equation}
F({\bm r},t)=\int^{t-{\bm n}\cdot {\bm r}/c}\dd\phi\left[-c\frac{eA(\phi)\cdot p}{n\cdot p}+c\frac{e^2A^2(\phi)}{2n\cdot p}\right],
\label{a6}
\end{equation}
which is {\it the relativistic Volkov phase}~\cite{Volkov}. Next, by accounting for Eq.~\eqref{a4}, Eq.~\eqref{a3} takes the form
\begin{align}
\biggl[\frac{1}{c^2}(\ii\partial_t&-V)^2-\hat{\bm p}^2-(m_{\rm e}c)^2-\frac{2}{c^2}F'(\ii\partial_t-V-E_{\bm p})\nonumber\\
&+2\left(e{\bm A}+\frac{1}{c}F'{\bm n}\right)\cdot (\hat{\bm p}-{\bm p})\biggr]\phi_{\bm p}({\bm r},t)=0,
\label{a7}
\end{align}
where one has to keep in mind that $F'$ satisfies Eq.~\eqref{a5}. In the next step, we take care of rest-mass oscillations by replacing
$\phi_{\bm p}({\bm r},t)$ by
\begin{equation}
\phi_{\bm p}({\bm r},t)=\ee^{-\ii m_{\rm e}c^2t}\,\bar\phi_{\bm p}({\bm r},t).
\label{a8}
\end{equation}
After the relativistic reduction, Eq.~\eqref{a7} can be conveniently rewritten in the form resembling the Schr\"odinger equation,
\begin{equation}
\left(\ii\partial_t-V-\frac{\hat{\bm p}^2}{2m_{\rm e}}+\hat{h}\right)\bar\phi_{\bm p}({\bm r},t)=0,
\label{a9}
\end{equation}
where the operator $\hat{h}$ has been introduced,
\begin{align}
\hat{h}&=\frac{1}{m_{\rm e}}\left(e{\bm A}+\frac{1}{c}F'{\bm n}\right)\cdot (\hat{\bm p}-{\bm p})\label{a10}\\
&+\frac{1}{2m_{\rm e}c^2}\left[(\ii\partial_t-V)^2-2F'(\ii\partial_t-V+m_{\rm e}c^2-E_{\bm p})\right].\nonumber
\end{align}
Note that $\hat{h}$ can be treated as a perturbation to the field-free nonrelativistic problem. Specifically, for highly energetic electrons, the first 
term in Eq.~\eqref{a10} represents a correction of the order of $(\hat{\bm p}-{\bm p})$ (see, Ref.~\cite{Faisal}). The second term, on the other hand, represents a relativistic 
correction of the order of $1/c^2$. In that case, expanding the wave function $\bar\phi_{\bm p}({\bm r},t)$ as
\begin{equation}
\bar\phi_{\bm p}({\bm r},t)=\bar\phi_{\bm p}^{(0)}({\bm r},t)+\bar\phi_{\bm p}^{(1)}({\bm r},t)+...,
\label{a11}
\end{equation}
we arrive at the corresponding lowest-order differential equations, 
\begin{align}
\left(\ii\partial_t-V-\frac{\hat{\bm p}^2}{2m_{\rm e}}\right)\bar\phi_{\bm p}^{(0)}({\bm r},t)&=0, \label{a12}\\
\left(\ii\partial_t-V-\frac{\hat{\bm p}^2}{2m_{\rm e}}\right)\bar\phi_{\bm p}^{(1)}({\bm r},t)&=-\hat{h}\bar\phi_{\bm p}^{(0)}({\bm r},t). \label{a13}
\end{align}
Hence, $\bar\phi_{\bm p}^{(0)}({\bm r},t)$ is the field-free solution of the nonrelativistic wave equation, whereas 
$\bar\phi_{\bm p}^{(1)}({\bm r},t)$ represents its first-order correction. Thus, for energetic electrons 
and as long as we are interested in leading nondipole corrections, the scattering wave function of the electron can be represented in a product form,
\begin{equation}
\psi_{\bm p}({\bm r},t)=\ee^{-\ii m_{\rm e}c^2t+\ii F({\bm r},t)}\bar{\phi}_{\bm p}^{(0)}({\bm r},t),
\label{a14}
\end{equation}
known as {\it the Coulomb-Volkov wave function}. We recall that $\bar{\phi}_{\bm p}^{(0)}({\bm r},t)$ is the scattering solution of the nonrelativistic 
Coulomb problem [Eq.~\eqref{a12}], whereas the Volkov phase $F({\bm r},t)$ in its full relativistic form is given by Eq.~\eqref{a6}. For completeness,
following Ref.~\cite{Berestetskii}, we write down that
\begin{align}
\bar{\phi}_{\bm p}^{(0)}({\bm r},t)&=\ee^{\pi\nu/2}\ee^{\ii{\bm p}\cdot{\bm r}-\ii{\bm p}^2t/(2m_{\rm e})}\Gamma(1-\ii\nu)\nonumber\\
&\times {}_1F_1(\ii\nu,1,\ii(|{\bm p}|\,|{\bm r}|-{\bm p}\cdot {\bm r})),
\label{a15}
\end{align}
where $\nu={\cal Z}/(a_0|{\bm p}|)$, whereas ${}_1F_1(a,b,z)$ is the confluent hypergeometric function of the first kind.

For our further purpose, we transform the Coulomb-Volkov wave into the length gauge. In that case, 
Eq.~\eqref{a14} becomes
\begin{equation}
\psi_{\bm p}^{(L)}({\bm r},t)=\ee^{-\ii m_{\rm e}c^2t-\ii e{\bm A}({\bm r},t)\cdot{\bm r}+\ii F({\bm r},t)}\bar{\phi}_{\bm p}^{(0)}({\bm r},t),
\label{a16}
\end{equation}
where we extract now the leading nondipole terms. Since the laser field depends on the retarded time
$t_{\rm R}=t-{\bm n}\cdot {\bm r}/c$, we account for the retardation correction by expanding the vector potential in powers of $1/c$ such that
\begin{equation}
{\bm A}({\bm r},t)={\bm A}(t-{\bm n}\cdot {\bm r}/c)\approx {\bm A}(t)+\frac{{\bm n}\cdot {\bm r}}{c} \,{\bm{\mathcal{E}}}(t),
\label{a17}
\end{equation}
where ${\bm{\mathcal{E}}}(t)=-\partial_t {\bm A}(t)$ is the electric field describing the laser light in the dipole approximation. 
Then, the gauge transformation phase factor in Eq.~\eqref{a16} can be replaced by
\begin{equation}
\ee^{-\ii e{\bm A}({\bm r},t)\cdot{\bm r}}\approx \ee^{-\ii e{\bm A}(t)\cdot{\bm r}}\left[1-\ii e{\bm{\mathcal{E}}}(t)
\cdot{\bm r}\,\frac{{\bm n}\cdot {\bm r}}{c}\right],
\label{a18}
\end{equation}
which we will relate to as {\it the gauge transformation correction}.
The second {\it retardation correction} follows from the integration limit in Eq.~\eqref{a6}. Hence, up to the first order in $1/c$,
\begin{align}
F({\bm r},t)&=\frac{c}{n\cdot p}\int^{t-{\bm n}\cdot {\bm r}/c}\dd\phi\left[e{\bm A}(\phi)\cdot{\bm p}-\frac{1}{2}e^2{\bm A}^2(\phi)\right]\nonumber\\
&\approx\frac{c}{n\cdot p}\int^{t}\dd\phi\left[e{\bm A}(\phi)\cdot{\bm p}-\frac{1}{2}e^2{\bm A}^2(\phi)\right]\nonumber\\
&-\frac{{\bm n}\cdot {\bm r}}{n\cdot p}\,\left[e{\bm A}(t)\cdot{\bm p}-\frac{1}{2}e^2{\bm A}^2(t)\right].
\label{a19}
\end{align}
Finally, we shall account for the correction originating from the term $1/(n\cdot p)$ present in the definition of $F({\bm r},t)$
above. In the leading order in $1/c$, we obtain that
\begin{equation}
\frac{1}{n\cdot p}\approx\frac{1}{m_{\rm e}c}\left(1+\frac{{\bm n}\cdot {\bm p}}{m_{\rm e}c}\right),
\label{a20}
\end{equation}
which is known as the {\it Nordsieck correction}~\cite{Nordsieck}. As it was argued in Ref.~\cite{KK1} in the context of ionization, 
this term can be interpreted as arising from an effective momentum-dependent mass,
\begin{equation}
m_{\rm eff}=m_{\rm e}-{\bm n}\cdot{\bm p}/c,
\label{effective}
\end{equation}
and is responsible for the electron recoil when interacting with the laser field. Thus, it can be attributed to the radiation pressure 
exerted on the electron by the field. Hence, combining Eqs.~\eqref{a19} and~\eqref{a20}, we obtain in the leading order in $1/c$ that the relativistic
Volkov phase can be approximated as
\begin{align}
F({\bm r},t)&\approx \frac{1}{m_{\rm e}}\left(1+\frac{{\bm n}\cdot {\bm p}}{m_{\rm e}c}\right)\int^{t}\dd\phi
\left[e{\bm A}(\phi)\cdot{\bm p}-\frac{1}{2}e^2{\bm A}^2(\phi)\right]\nonumber\\
&-\frac{{\bm n}\cdot {\bm r}}{m_{\rm e}c}\left[e{\bm A}(t)\cdot{\bm p}-\frac{1}{2}e^2{\bm A}^2(t)\right].
\label{a21}
\end{align}
This together with Eq.~\eqref{a15} define the scattering wave function of the initial electron [Eq.~\eqref{a14}]. It will be used next
to derive the formula for the probability amplitude of LARR beyond the dipole approximation.

\subsection{Probability amplitude of LARR}
\label{amplitude}

\subsubsection{Monochromatic electron wave}
\label{monochromatic}

Consider an electron of momentum ${\bm p}$ recombining to the ground state of a hydrogen-like atom with the atomic number ${\cal Z}$.
Starting with the Klein-Gordon equation, one can show through the relativistic reduction that, in the leading order in $1/c$, for the final electron state
one can take 
\begin{equation}
\psi_B({\bm r},t)=\ee^{-\ii m_{\rm e} c^2t-\ii E_Bt}\psi_B({\bm r}),
\label{a22}
\end{equation}
where
\begin{equation}
\psi_B({\bm r})=\frac{1}{\sqrt{\pi}}\biggl(\frac{\cal Z}{a_0}\biggr)^{3/2}\ee^{-{\cal Z}r/a_0},
\label{a23}
\end{equation}
whereas for the bounding energy, $E_B=-{\cal Z}^2E_0/2$. Similarly, one can identify in the leading order, that the interaction Hamiltonian in the length gauge is
\begin{equation}
\hat{H}_I({\bm r},t)=-e\hat{\bm{\mathcal{E}}}_{\bm K}(\hat{\bm r},t)\cdot\hat{\bm r},
\label{a24}
\end{equation}
where the quantum field operator $\hat{\bm{\mathcal{E}}}_{\bm K}(\hat{\bm r},t)$ represents the radiated photon. More specifically, 
\begin{equation}
\hat{\bm{\mathcal E}}_{\bm K}(\hat{\bm r},t)=\hat{\bm{\mathcal E}}_{\bm K}^{(-)}(\hat{\bm r},t)+h.c.,
\label{a25}
\end{equation}
with
\begin{align}
\hat{\bm{\mathcal E}}_{\bm K}^{(-)}(\hat{\bm r},t)=-\ii{\bm \varepsilon}_{\bm K}\sqrt{\frac{\omega_{\bm K}}{2\epsilon_0 V}}\,\hat{a}_{\bm K}^\dag
\ee^{\ii (\omega_{\bm K}t-{\bm K}\cdot\hat{\bm r})},
\label{a26}
\end{align}
where $V$ is the quantization volume and $\hat{a}_{\bm K}^\dagger$ is the creation operator. Note that in Eq.~\eqref{a25}
there is also the hermitian conjugate of $\hat{\bm{\mathcal E}}_{\bm K}^{(-)}(\hat{\bm r},t)$.

Substituting now Eqs.~\eqref{a16},~\eqref{a22}, and~\eqref{a24} into the definition of the probability amplitude of LARR [Eq.~\eqref{a1}], we obtain
\begin{align}
{\cal A}({\bm p})&=e\sqrt{\frac{\omega_{\bm K}}{2\varepsilon_0V}}\int_{-\infty}^\infty\dd t\int\dd^3r\,\psi_B^*({\bm r})({\bm\varepsilon}_{\bm K}
\cdot{\bm r})\psi_{\bm p}^{(0)}({\bm r},t)\nonumber\\
&\times \ee^{\ii(E_B+\omega_{\bm K})t-\ii[{\bm K}+e{\bm A}({\bm r},t)]\cdot {\bm r}-\ii F({\bm r},t)},
\label{a27}
\end{align}
where $F({\bm r},t)$ is given by Eq.~\eqref{a21}, whereas $\psi_{\bm p}^{(0)}({\bm r},t)$ by Eq.~\eqref{a15}. Taking into account those definitions,
we introduce now the following abbreviations,
\begin{align}
Q&=E_B+\omega_{\bm K}-\frac{{\bm p}^2}{2m_{\rm e}},\label{a28}\\
{\bm q}(t)&={\bm p}-e{\bm A}(t)-{\bm K}\nonumber\\
&-\frac{{\bm n}}{m_{\rm e}c}\bigl[e{\bm A}(t)\cdot{\bm p}-\frac{1}{2}e^2{\bm A}^2(t)\bigr],\label{a29}\\
H(t)&=\frac{1}{m_{\rm e}}\Bigl(1+\frac{{\bm n}\cdot{\bm p}}{m_{\rm e}c}\Bigr)\nonumber\\
&\times\int^t\dd\tau\bigl[e{\bm A}(\tau)\cdot{\bm p}-\frac{1}{2}e^2{\bm A}^2(\tau)\bigr].\label{a30}
\end{align}
Then, the LARR probability amplitude which includes nondipole first-order corrections~\eqref{a27} takes the form,
\begin{align}
{\cal A}({\bm p})&=\frac{{\cal Z}e}{Va_0}\sqrt{\frac{{\cal Z}\omega_{\bm K}}{2\pi\varepsilon_0a_0}}\ee^{\pi\nu/2}\Gamma(1-\ii\nu)\int_{-\infty}^\infty\dd t\,\ee^{\ii Qt+\ii H(t)}\nonumber\\
&\times\int\dd^3r\,({\bm\varepsilon}_{\bm K}\cdot{\bm r})\Bigl[1-\ii \left(e{\bm{\mathcal{E}}}(t)\cdot{\bm r}\right)
\frac{{\bm n}\cdot {\bm r}}{c}\Bigr]\nonumber\\
&\times{}_1F_1(\ii\nu,1,\ii(|{\bm p}|\,|{\bm r}|-{\bm p}\cdot {\bm r}))\,\ee^{-{\cal Z}r/a_0+\ii{\bm q}(t)\cdot {\bm r}},\label{a31}
\end{align}
where we have inserted the scattering Coulomb state [Eq.~\eqref{a15}] and the bound state wave function [Eq.~\eqref{a23}].

Note that Eq.~\eqref{a31} can be represented as
\begin{equation}
{\cal A}({\bm p})={\cal N}\int_{-\infty}^\infty\dd t\,\ee^{\ii Qt+\ii H(t)}\Bigl[{\cal B}(t)-\frac{1}{c}{\cal C}(t)\Bigr],
\label{a33}
\end{equation}
where ${\cal N}=\frac{{\cal Z}e}{Va_0}\sqrt{\frac{{\cal Z}\omega_{\bm K}}{2\pi\varepsilon_0a_0}}\ee^{\pi\nu/2}\Gamma(1-\ii\nu)$ and where we have introduced
the following functions,
\begin{align}
{\cal B}(t)&=\int\dd^3r\,({\bm\varepsilon}_{\bm K}\cdot{\bm r})\ee^{-{\cal Z}r/a_0+\ii{\bm q}(t)\cdot {\bm r}}\nonumber\\
&\times{}_1F_1(\ii\nu,1,\ii(|{\bm p}|\,|{\bm r}|-{\bm p}\cdot {\bm r})),\label{a34}\\
{\cal C}(t)&=\int\dd^3r\,({\bm\varepsilon}_{\bm K}\cdot{\bm r})(e\bm{\mathcal{E}}(t)\cdot{\bm r})({\bm n}\cdot{\bm r})
\ee^{-{\cal Z}r/a_0+\ii{\bm q}(t)\cdot {\bm r}}\nonumber\\
&\times{}_1F_1(\ii\nu,1,\ii(|{\bm p}|\,|{\bm r}|-{\bm p}\cdot {\bm r}))\label{a35}.
\end{align}
Each of these functions can be calculated with the help of the so-called {\it Nordsieck integral}, which is defined as
\begin{equation}
f(\nu,\lambda,{\bm q},{\bm p})=\int\dd^3r\frac{\ee^{-\lambda r}}{r}\ee^{\ii{\bm q}\cdot {\bm r}}{}_1F_1(\ii\nu,1,\ii(|{\bm p}|\,|{\bm r}|
-{\bm p}\cdot {\bm r})).
\label{a36}
\end{equation}
Because the above integral can be solved analytically, leading to
\begin{equation}
f(\nu,\lambda,{\bm q},{\bm p})=4\pi\zeta(1+\xi)^{-\ii\nu},
\label{a37}
\end{equation}
with
\begin{equation}
\zeta=\frac{1}{\lambda^2+{\bm q}^2},\quad \xi=-2\zeta\left({\bm p}\cdot{\bm q}+\ii \lambda |{\bm p}|\right),
\label{a38}
\end{equation}
one can derive the explicit formulas for the functions~\eqref{a34} and~\eqref{a35}. This is done using the following relations,
\begin{align}
{\cal B}(t)&=\ii\frac{\partial}{\partial\lambda}({\bm\varepsilon}_{\bm K}\cdot {\bm\nabla}_{\bm q})f(\nu,\lambda,{\bm q},{\bm p}),\label{a39}\\
{\cal C}(t)&=-\ii\frac{\partial}{\partial\lambda}({\bm\varepsilon}_{\bm K}\cdot {\bm\nabla}_{\bm q})(e\bm{\mathcal{E}}(t)\cdot {\bm\nabla}_{\bm q})
({\bm n}\cdot{\bm\nabla}_{\bm q})f(\nu,\lambda,{\bm q},{\bm p}),\label{a40}
\end{align}
where $\lambda={\cal Z}/a_0$ and ${\bm q}={\bm q}(t)$ is given by Eq.~\eqref{a29}.

Let us consider now the radiative recombination which is assisted by a short laser pulse, lasting from 0 to $T_p$. In that case, the probability amplitude
of LARR~\eqref{a33} is divergent due to the presence of term ${\cal B}(t)$. However, as we have elaborated in Ref.~\cite{Deeksha}, such kind of divergence can
be treated using the Boca-Florescu transformation~\cite{Boca}. Hence, following the procedure described in Appendix~\ref{app}, the LARR probability amplitude~\eqref{a33} becomes
\begin{align}
{\cal A}({\bm p})&={\cal N}\biggl\{2\pi\delta(Q){\cal B}(0)\ee^{\ii H(T_p)/2}\cos(H(T_p)/2)\nonumber\\
&+\ii {\cal P}\left(\frac{1}{Q}\right)\int_0^{T_p}\dd t\,\ee^{\ii Qt+\ii H(t)}\Bigl[\dot{\cal B}(t)+\ii \dot{H}(t){\cal B}(t)\Bigr]\nonumber\\
&-\frac{1}{c}\int_0^{T_p}\dd t\,\ee^{\ii Qt+\ii H(t)}{\cal C}(t)\biggr\},
\label{a41}
\end{align}
where the dot stands for the time derivative.
Note that, compared to Eq.~\eqref{a33}, we have restricted the integration limit in the last integral as ${\cal C}(t)$ takes nonzero values only
for $t\in[0,T_p]$. 


At this point, it is worth noting that the probability amplitude of LARR~\eqref{a41} contains a contribution from the field-free
process. This follows from the presence of the Dirac $\delta$ function, which determines the energy conservation condition, $Q=0$ [Eq.~\eqref{a28}]. 
Based on this condition, we expect to observe a peak in the spectrum of LARR at energy,
\begin{equation}
\omega_{\bm K}=\frac{{\bm p}^2}{2m_{\rm e}}+|E_B|.
\label{a42}
\end{equation}
The appearance of this peak was also predicted in Ref.~\cite{Deeksha} for the laser-assisted radiative attachment, in the framework based on the dipole
approximation. Hence, it appears as a rather universal feature. As we have already argued in Ref.~\cite{Deeksha}, the reason for the field-free peak
is that laser-assisted processes occur even in the absence of the laser field. Thus, the corresponding field-free spectrum is embedded in the laser-induced
one. As one can see from Eq.~\eqref{a41}, there is a significant contribution from the latter, corresponding to $Q\neq 0$. Thus, we predict to observe
a broad spectrum of the induced radiation along with the peak at $Q=0$.

For our further purposes, we rewrite Eq.~\eqref{a41} in a more transparent way. Namely,
\begin{equation}
{\cal A}({\bm p})={\cal N}{\cal R}({\bm p}),
\label{a422}
\end{equation}
where
\begin{equation}
{\cal R}({\bm p})={\cal R}_0({\bm p})\delta(Q)+{\cal R}_1({\bm p}){\cal P}\left(\frac{1}{Q}\right)+{\cal R}_2({\bm p}),
\label{a43}
\end{equation}
with the partial LARR contributions defined as
\begin{align}
{\cal R}_0({\bm p})&=2\pi{\cal B}(0)\ee^{\ii H(T_p)/2}\cos(H(T_p)/2),\label{a44}\\
{\cal R}_1({\bm p})&=\ii\int_0^{T_p}\dd t\,\ee^{\ii Qt+\ii H(t)}\Bigl[\dot{\cal B}(t)+\ii \dot{H}(t){\cal B}(t)\Bigr],\label{a45}\\
{\cal R}_2({\bm p})&=-\frac{1}{c}\int_0^{T_p}\dd t\,\ee^{\ii Qt+\ii H(t)}{\cal C}(t).\label{a46}
\end{align}
As one can see, in order to calculate the probability amplitude~\eqref{a422}, we need to perform the remaining integrals in Eqs.~\eqref{a45}
and~\eqref{a46}. In actual computations, we define time-dependent functions,
\begin{align}
{\cal R}_1({\bm p},t)&=\ii\int_0^{t}\dd \tau\,\ee^{\ii Q\tau+\ii H(\tau)}\Bigl[\dot{\cal B}(\tau)+\ii \dot{H}(\tau){\cal B}(\tau)\Bigr],\label{a45bis}\\
{\cal R}_2({\bm p},t)&=-\frac{1}{c}\int_0^{t}\dd \tau\,\ee^{\ii Q\tau+\ii H(\tau)}{\cal C}(\tau).\label{a46bis}
\end{align}
Then, we solve the following system of differential equations, 
\begin{equation}
\label{a47new}
\left\{\begin{array}{rcl}\displaystyle
\dot{\bm A}(t)&=&-{\bm{\mathcal E}}(t),\\
\dot{H}(t)&=&\frac{1}{m_{\rm e}}\Bigl(1+\frac{{\bm n}\cdot{\bm p}}{m_{\rm e}c}\Bigr)\bigl[e{\bm A}(t)\cdot{\bm p}-\frac{1}{2}e^2{\bm A}^2(t)\bigr],\\
\dot{\cal R}_1({\bm p},t)&=&\ii\ee^{\ii Qt+\ii H(t)}\left[\dot{\cal B}(t)+\ii \dot{H}(t){\cal B}(t)\right],\\
\dot{\cal R}_2({\bm p},t)&=&-\frac{1}{c}\ee^{\ii Qt+\ii H(t)}{\cal C}(t),
\end{array} \right.
\end{equation}
which we integrate from $t=0$ to $t=T_p$. This is provided that we know the explicit form of the electric field, ${\bm{\mathcal E}}(t)$.
As a result, we obtain the values of $H(T_p)$, ${\cal R}_1({\bm p},T_p)={\cal R}_1({\bm p})$, and ${\cal R}_2({\bm p},T_p)={\cal R}_2({\bm p})$,
which allow us to calculate the probability amplitude of LARR, given by Eq.~\eqref{a422}.

In deriving the above formulas, we have assumed a monochromatic electron wave $\psi_{\bm p}({\bm r},t)$ [Eq.~\eqref{a14}] impinging on an ion in the presence of a laser pulsed field. This has led
to appearance of singular distributions $\delta(Q)$ and ${\cal P}(\frac{1}{Q})$ in Eq.~\eqref{a43}. As it follows from our previous work~\cite{Deeksha}, those singularities
can be removed provided that, instead of a monochromatic electron wave, we take a coherent superposition of such waves. This is elaborated on in the next section.

\subsubsection{Coherent electron wave packet}
\label{coherent}

Now we consider a coherent electron wave packet that recombines with an ion in the presence of a laser pulsed field. The initial electron wave packet 
$\psi_{\bm p}[{\bm r},t|f_{\bm p}]$ is defined as a weighted superposition of monochromatic electron waves $\psi_{\bm p}({\bm r},t)$ [Eq.~\eqref{a14}],
\begin{equation}
\psi_{\bm p}[{\bm r},t|f_{\bm p}]=\int\dd^3{\bm \kappa}\,\psi_{\bm \kappa}({\bm r},t)f_{\bm p}({\bm \kappa}).
\label{a47}
\end{equation}
Here, we assume that the wave packet profile $f_{\bm p}({\bm \kappa})$ is such that
\begin{equation}
f_{\bm p}({\bm \kappa})\approx \delta^{(3)}({\bm \kappa}-{\bm p}),
\label{a48}
\end{equation}
i.e., it describes a nearly monochromatic electron beam.
Consequently, the LARR probability amplitude~\eqref{a27} integrated over the initial electron momentum profile equals
\begin{equation}
\langle {\cal A}({\bm p})\rangle=\int\dd^3{\bm \kappa}\,{\cal A}({\bm \kappa})f_{\bm p}({\bm \kappa}).
\label{a49}
\end{equation}
Substituting here Eq.~\eqref{a422}, we obtain
\begin{align}
\langle{\cal A}({\bm p})\rangle&={\cal N}\langle{\cal R}({\bm p})\rangle\label{a50}\\
&={\cal N}\biggl[{\cal R}_0({\bm p})\langle\delta(Q)\rangle+{\cal R}_1({\bm p})\left\langle{\cal P}
\left(\frac{1}{Q}\right)\right\rangle+{\cal R}_2({\bm p})\biggr],\nonumber
\end{align}
which implicitly defines $\langle{\cal R}({\bm p})\rangle$ and
where each contribution ${\cal N}{\cal R}_j({\bm p})$ (for $j=0,1,2$) [Eqs.~\eqref{a44},~\eqref{a45}, and~\eqref{a46}] is slowly 
varying function of ${\bm p}$. Note that this is in contrast to the singular distributions $\delta(Q)$ and ${\cal P}(\frac{1}{Q})$,
which are averaged in Eq.~\eqref{a50} over the initial electron momentum spread. In doing so, we define
\begin{equation}
\left\langle\frac{1}{Q+\ii\epsilon}\right\rangle=\int\dd^3{\bm \kappa}\,\frac{f_{\bm p}({\bm \kappa})}{Q_{\bm \kappa}+\ii\epsilon},\label{a51}
\end{equation}
with $Q_{\bm \kappa}=E_B+\omega_{\bm K}-{\bm \kappa}^2/(2m_{\rm e})$. Following the Sokhotski-Plemelj formula~\eqref{b88}, this allows us to calculate
\begin{align}
\langle\delta(Q)\rangle&=-\frac{1}{\pi}\lim_{\varepsilon\rightarrow 0^+}{\rm Im}\left\langle\frac{1}{Q+\ii\varepsilon}\right\rangle,\label{a52}\\
\left\langle {\cal P}\left(\frac{1}{Q}\right)\right\rangle&=\lim_{\varepsilon\rightarrow 0^+}{\rm Re}\left\langle\frac{1}{Q+\ii\epsilon}\right\rangle.
\label{a53}
\end{align}
Thus, the field-free singularity at $Q=0$ is smeared out by the initial electron momentum profile.
Note also that Eq.~\eqref{a50} accounts coherently for LARR contributions arising from different wave components of the initial electron wave packet.

As an example, we shall assume that the initial electron wave packet is well-collimated such that it is represented by the following
momentum distribution,
\begin{equation}
f_{\bm p}({\bm \kappa})\approx \frac{1}{\pi}\frac{\Delta |{\bm p}|}{(\kappa_\|-|{\bm p}|)^2+(\Delta |{\bm p}|)^2}\delta^{(2)}({\bm \kappa}_\perp).
\label{a59}
\end{equation}
Here, the contributing longitudinal $\kappa_\|$ and transverse ${\bm \kappa}_\perp$ momentum components are defined with respect to the direction of the central 
electron momentum ${\bm p}$,
\begin{equation}
\kappa_\|=\frac{{\bm \kappa}\cdot{\bm p}}{|{\bm p}|},\qquad {\bm \kappa}_\perp={\bm \kappa}-\kappa_\|\frac{{\bm p}}{|{\bm p}|}.
\label{a60}
\end{equation}
Thus, while we neglect the electron momentum spread in the transverse direction, we assume the Lorentzian profile of the longitudinal 
momenta forming the initial electron wave packet. Note that the latter is centered at $\kappa_\|=|{\bm p}|$ and has the half-width at half maximum 
(HWHM) $\Delta |{\bm p}|$. Hence, the average~\eqref{a51}
equals
\begin{equation}
\left\langle\frac{1}{Q+\ii\varepsilon}\right\rangle=\frac{1}{E_B+\omega_{\bm K}-\frac{{\bm p}^2}{2m_{\rm e}}+\ii\frac{\kappa_0}{m_{\rm e}}
\Delta |{\bm p}|},
\label{a61}
\end{equation}
where $\kappa_0=\sqrt{2m_{\rm e}(E_B+\omega_{\bm K})}$ and where the limit $\varepsilon\rightarrow 0^+$ was taken. Following Eqs.~\eqref{a52}
and~\eqref{a53}, we obtain
\begin{align}
\langle\delta(Q)\rangle&=\frac{1}{\pi}\frac{\frac{\kappa_0}{m_{\rm e}}
\Delta |{\bm p}|}{(E_B+\omega_{\bm K}-\frac{{\bm p}^2}{2m_{\rm e}})^2+(\frac{\kappa_0}{m_{\rm e}}
\Delta |{\bm p}|)^2},\label{a62}\\
\left\langle {\cal P}\left(\frac{1}{Q}\right)\right\rangle&=\frac{E_B+\omega_{\bm K}-\frac{{\bm p}^2}{2m_{\rm e}}}{(E_B+\omega_{\bm K}
-\frac{{\bm p}^2}{2m_{\rm e}})^2+(\frac{\kappa_0}{m_{\rm e}}
\Delta |{\bm p}|)^2},
\label{a63}
\end{align}
which determine the probability amplitude of LARR integrated over the initial electron momentum distribution, Eq.~\eqref{a50}.
Keeping this in mind, we shall define below the corresponding energy distribution of LARR.

\subsection{Energy probability distribution of LARR}
\label{energy_distr}

The total energy [per the initial electron flux $j_{\rm e}({\bm p})$] irradiated due to the interaction 
of an electron wave packet with an ion in the presence of a pulsed laser field equals
\begin{equation}
E_{\bm K}({\bm p})=\frac{1}{j_{\rm e}({\bm p})}\int\frac{V\dd^3{\bm K}}{(2\pi)^3}\,\omega_{\bm K} |\langle{\cal A}({\bm p})\rangle|^2,
\label{a54}
\end{equation}
where $\langle{\cal A}({\bm p})\rangle$ is given by Eq.~\eqref{a50}. Here, the integration is over the density of final radiation states, 
$V\dd^3{\bm K}/(2\pi)^3$. Moreover, using the relation $|{\bm K}|=\omega_{\bm K}/c$ we derive that 
$\dd^3{\bm K}=\omega_{\bm K}^2\dd\omega_{\bm K}\dd^2\Omega_{\bm K}/c^3$, where $\dd^2\Omega_{\bm K}$ is the solid angle of emitted photons. 
The initial electron flux is
\begin{equation}
j_{\rm e}({\bm p})=\frac{|{\bm p}|}{m_{\rm e}}\frac{1}{V}.
\label{a55}
\end{equation}
Hence, we obtain
\begin{align}
E_{\bm K}({\bm p})&=\frac{\nu^4\ee^{\pi\nu}}{\sinh(\pi\nu)}\frac{\alpha m_{\rm e}|{\bm p}|^2}{(2\pi)^2c^2}\nonumber\\
&\times\int\dd^2\Omega_{\bm K}\int\dd\omega_{\bm K}\,\omega_{\bm K}^4|\langle {\cal R}({\bm p})\rangle|^2,
\label{a56}
\end{align}
where we have used the following property: $|\Gamma(1-\ii\nu)|^2=\pi\nu/\sinh(\pi\nu)$.
Rewriting this formula as
\begin{equation}
E_{\bm K}({\bm p})=\int\dd^2\Omega_{\bm K}\int\dd\omega_{\bm K}\,\frac{\dd^3E_{\bm K}({\bm p})}{\dd\omega_{\bm K}\dd^2\Omega_{\bm K}},
\label{a57}
\end{equation}
we conclude that the triply differential energy distribution (per the initial electron flux) of photons emitted
in the solid angle $\dd^2\Omega_{\bm K}$ and having energy within the interval $(\omega_{\bm K},\omega_{\bm K}+\dd\omega_{\bm K})$ equals
\begin{equation}
\frac{\dd^3E_{\bm K}({\bm p})}{\dd\omega_{\bm K}\dd^2\Omega_{\bm K}}=\frac{\nu^4\ee^{\pi\nu}}{\sinh(\pi\nu)}\frac{\alpha m_{\rm e}|{\bm p}|^2}{(2\pi)^2c^2}\,\omega_{\bm K}^4|\langle {\cal R}({\bm p})\rangle|^2,
\label{a58}
\end{equation}
where $\langle{\cal R}({\bm p})\rangle$ has been defined in Eq.~\eqref{a50}. 

While the above derivations are quite general, for numerical illustrations we need to specify certain models of the initial electron
wave packet and the laser field. This is done in the following section, along with an analysis of properties of emitted radiation.

\section{Numerical illustrations}
\label{numerics}

Following the theoretical framework introduced above, we are going to analyze in this section possibilities for either extending the range
of the LARR energy spectrum or enhancing it; both being of potential practical interest. In doing so, we shall use tailored laser pulses 
which allow for a great control over LARR energy and angular distributions. In this context, we shall investigate nondipole effects in
LARR along with a control of LARR radiation by chirping the assisting laser pulse.

\subsection{Nondipole effects in LARR}
\label{nondipole}

\begin{figure}
\begin{center}
\includegraphics[width=0.7\linewidth]{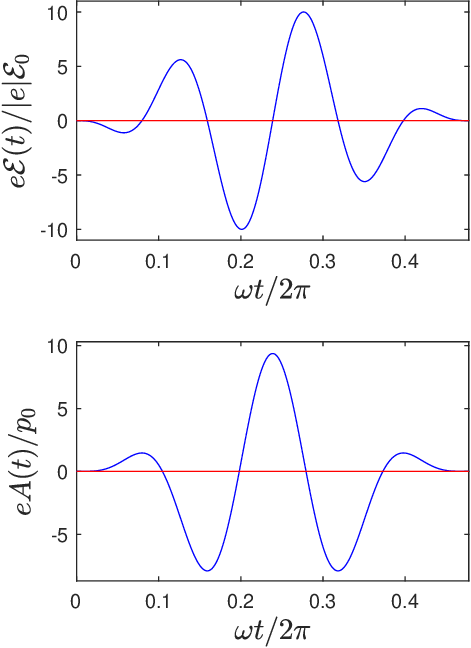}
\end{center}
\caption{Electric field ${\cal E}(t)$ and vector potential $A(t)$ of a three cycle ($N_{\rm osc}=3$) laser pulse
with a sine-squared envelope [Eq.~\eqref{n1}] such that $\omega=30.1~{\rm eV}=1.14E_0\,(\lambda=40~{\rm nm})$ and ${\cal E}=10{\cal E}_0$.
Atomic units are used.}
\label{fig1}
\end{figure}

Let us consider LARR in the presence of an $N_{\rm osc}$ cycle laser pulse that is linearly polarized along 
the $x$-axis. It is represented by the electric field vector $\bm{\mathcal{E}}(t)={\mathcal{E}}(t)\bm{e}_{x}$ with the sine-squared envelope,
\begin{align}
{\mathcal{E}}(t)=\left\{ \begin{array}{ll}\mathcal{E}\mathcal{N}_0\sin^{2}\left(\frac{\omega t}{2N_{\rm osc}}\right)\sin(\omega t),& 
\mbox{if\, $0\leqslant t\leqslant T_p$},\\
0, & \mbox{otherwise}.\end{array} \right.
\label{n1}
\end{align}
Here, $\omega$ is the carrier frequency and so the pulse duration equals $T_p=2\pi N_{\rm osc}/\omega$. Moreover, the scaling factor
$\mathcal{N}_0$ is adjusted such that 
\begin{equation}
\underset{0\leqslant t\leqslant T_p}{\rm max}|{\cal E}(t)|={\cal E}, 
\label{n2}
\end{equation}
meaning that ${\cal E}$ determines the peak amplitude of the electric field~\eqref{n1}. For numerical illustrations, we choose 
$\omega=1.14E_0\,(\lambda=40\,{\rm nm}), \mathcal{E}=10 \mathcal{E}_{0}$, and $N_{\rm osc}=3$. Thus, in Fig.~\ref{fig1}, we present 
the time-dependence of the electric field ${\cal E}(t)$ (upper panel) and the corresponding vector potential 
$A(t)=-\int_0^t\dd\tau\,{\cal E}(\tau)$ (lower panel). This will be useful when analyzing the emitted LARR radiation.

\begin{figure}
\centering
  \includegraphics[width=0.8\linewidth]{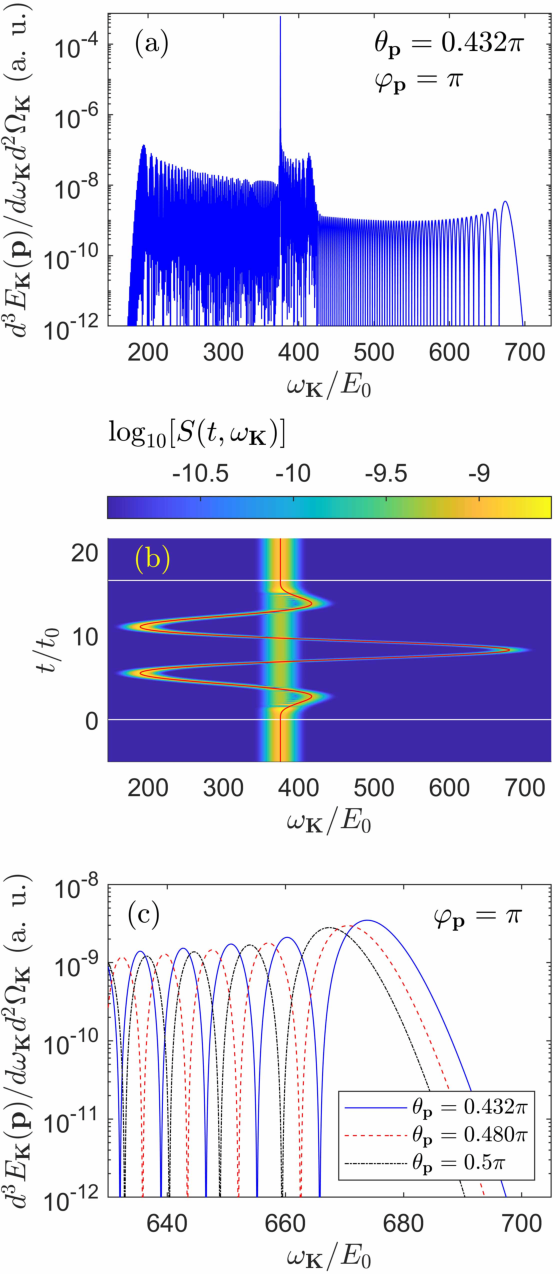}
\caption{(a) LARR energy spectrum emitted when the electron recombines with a hydrogen-like ion (${\cal Z}=4$) in the presence of a pulse
shown in Fig.~\ref{fig1}. While the pulse propagates along the $z$-axis and is linearly polarized in the $x$-direction, the electron
wave packet propagates in the $xz$-plane at a polar angle $\theta_{\bm p}=0.432\pi$. Its central momentum $|{\bm p}|$ corresponds to
the kinetic energy $E_{\bm p}=10$~keV, whereas the energy spread is $\Delta |{\bm p}|=2.74\times 10^{-5}p_0$ (for more details on the model of electron wave packet
see Sec.~\ref{coherent}). Also, the LARR radiation is linearly polarized along the $x$-axis while being emitted in the $z$-direction.
The presented results have been calculated based on the theory accounting for nondipole corrections discussed in Sec.~\ref{monochromatic}.
(b) Spectrogram of the data plotted in panel (a) (yellow ribbon) that was calculated using the Gaussian window [Eq.~\eqref{c4}] with the 
parameters: $\xi_T=0.1, \xi_W=0.03, \omega_1=-5E_0$, and $\omega_2=21.55E_0$. For comparison, Eq.~\eqref{n3} representing the energy emitted by an electron of momentum ${\bm p}$ 
that evolves in a laser pulse and is captured at time $t$ by the atom is plotted (red line). (c) LARR energy 
spectra near the cutoff for different polar angles of incident electron $\theta_{\bm p}$, as given in the legend.}
\label{fig2}
\end{figure}

In the following, we consider the laser pulse~\eqref{n1} that propagates in the $z$-direction, ${\bm n}={\bm e}_z$. Thus, in order
to observe the most pronounced radiation pressure effects, we shall investigate the emission of LARR in that direction,
${\bm n}_{\bm K}={\bm e}_z$. Moreover, we assume that the pulse is linearly polarized along the $x$-axis, ${\bm\varepsilon}_{\bm K}={\bm e}_x$. 
Regarding the electron wave packet [Eq.~\eqref{a59}], we shall center it at momentum
$|{\bm p}|$ such that $E_{\bm p}=\frac{{\bm p}^2}{2m_{\rm e}}=10$~keV, with the longitudinal spread characterized by the parameter
$\Delta |{\bm p}|=\Delta\sqrt{2m_{\rm e}E_{\bm p}}=10^{-6}\sqrt{2m_{\rm e}E_{\bm p}}=2.74\times 10^{-5}p_0$. In addition, in Figs.~\ref{fig2}(a) and~\ref{fig2}(b) the electron beam is propagating in the direction determined by 
the polar and azimuthal angles, $\theta_{\bm p}=0.432\pi$ and $\varphi_{\bm p}=\pi$, respectively, to collide with a positive ion
of ${\cal Z}=4$. This results in the energy spectrum presented in Fig.~\ref{fig2}(a). Let us stress that those results were calculated
based on Eq.~\eqref{a58}, with all nondipole corrections accounted for. As one can see, the spectrum consists of three plateaux,
having different extensions and heights, and a single peak at energy of roughly $\omega_{\bm K}=375E_0$. The location of the latter
agrees with Eq.~\eqref{a41}, meaning that it originates from the field-free process. The remaining properties of the spectrum can be estimated
based on the saddle-point approximation of the LARR probability amplitude~\eqref{a50}.

It follows from the saddle-point analysis that the probability amplitude~\eqref{a50} takes most significant values for
\begin{equation}
\omega_{\bm K}(t)=\omega_{\bm K}^{(0)}(t)-\frac{\bm{n}\cdot\bm{p}}{m_{e}^{2}c}\Big[e\bm{A}(t)\cdot\bm{p}-\frac{1}{2}e^2\bm{A}^{2}(t)\Big],
\label{n3}
\end{equation}
where
\begin{equation}
\omega_{\bm K}^{(0)}(t)=\frac{1}{2m_{\rm e}}[{\bm p}-e{\bm A}(t)]^2-E_B
\label{n4}
\end{equation}
is the saddle-point solution arising within the dipole approximation~\cite{Deeksha} (see, also Refs.~\cite{Milo1,Milo5,Milo6,Milo7,Tutmic}).
Thus, Eq.~\eqref{n3} represents in the leading nondipole order in $1/c$ the energy of photons emitted by the electron of momentum ${\bm p}$ that evolves 
in a laser pulse and recombines at time $t$. While Eq.~\eqref{n3} concerns recombination by a long-range Coulomb potential center --
in contrast, Eq.~\eqref{n4} has been derived for a short-range potential~\cite{Deeksha}. This indicates that, qualitatively, 
the Coulomb interaction does not have a significant influence on properties of emitted radiation. This has been also confirmed 
by our numerical results. Note that the second term in Eq.~\eqref{n3} represents the leading order nondipole correction to 
$\omega_{\bm K}^{(0)}(t)$. Based on the analysis presented in Sec.~\ref{VC}, one can recognize that this correction originates from the electron
recoil off a laser pulse. This is also supported by classical considerations in Appendix~\ref{classical}, where we analyze the electron dynamics in an electromagnetic field and derive the formula representing the electron kinetic energy beyond
the dipole approximation [see, Eq.~\eqref{ddd8} and the following discussion]. Therefore it is anticipated that, out of all $1/c$ corrections, it is the recoil correction that makes
the most pronounced contribution to the LARR spectrum. In order to confirm that, in Fig.~\ref{fig2}(b) we plot $\omega_{\bm K}(t)$
given by Eq.~\eqref{n3} (red line) against a spectrogram $S(t,\omega_{\bm K})$ of the signal presented in 
Fig.~\ref{fig2}(a) (yellow trace). For details on how to calculate the spectrogram, we refer the reader to Appendix~\ref{spect}.
Note that there are two horizontal lines as well plotted to mark the times when the laser pulse is turned on and off. It is easy to see
then that for as long as there is no laser field, only radiation with a central energy of roughly 375$E_0$ is generated. This continues 
even in the presence of the field, but less efficiently. On top of the field-free vertical line in the spectrogram, there is also a zig-zag pattern
which follows closely the analytical form of $\omega_{\bm K}(t)$ given by Eq.~\eqref{n3}. It obviously determines the extension of the 
LARR energy spectrum. Also, it is important to realize that the time-integral defining the LARR probability amplitude can be approximated 
by contributions from saddle points. As it follows from panel (b), the radiation of a given energy can be emitted at specific times 
(i.e., saddle points) during the pulse duration. This implicates that the high-energy plateau results from interference of 
two probability amplitudes originating from saddle points. As such, it exhibits regular and smooth oscillations. This is in contrast to 
the remaining plateaux. They originate from interference of four (the low-energy plateau) and six (the intermediate-energy plateau) 
saddle-point contributions, and so they exhibit more complex behavior. 

\begin{figure}
\centering
  \includegraphics[width=0.95\linewidth]{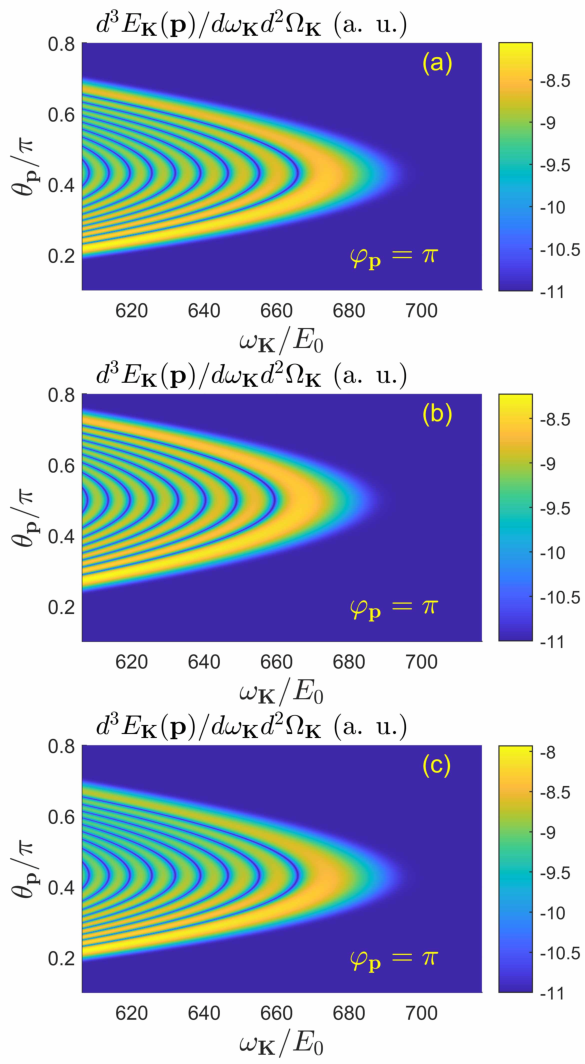}
\caption{Color mappings of the energy distributions of emitted LARR radiation near the cutoff at varied polar angle of recombining
electron $\theta_{\bm p}$, for the same parameters as in Fig.~\ref{fig2}.
While the results presented in panel (a) have been calculated by accounting for all nondipole corrections of the order $1/c$, panels (b)
and (c) correspond to the theory including either the retardation or the recoil correction, respectively.}
\label{fig3}
\end{figure}

\begin{figure*}
\begin{center}
\includegraphics[width=0.95\linewidth]{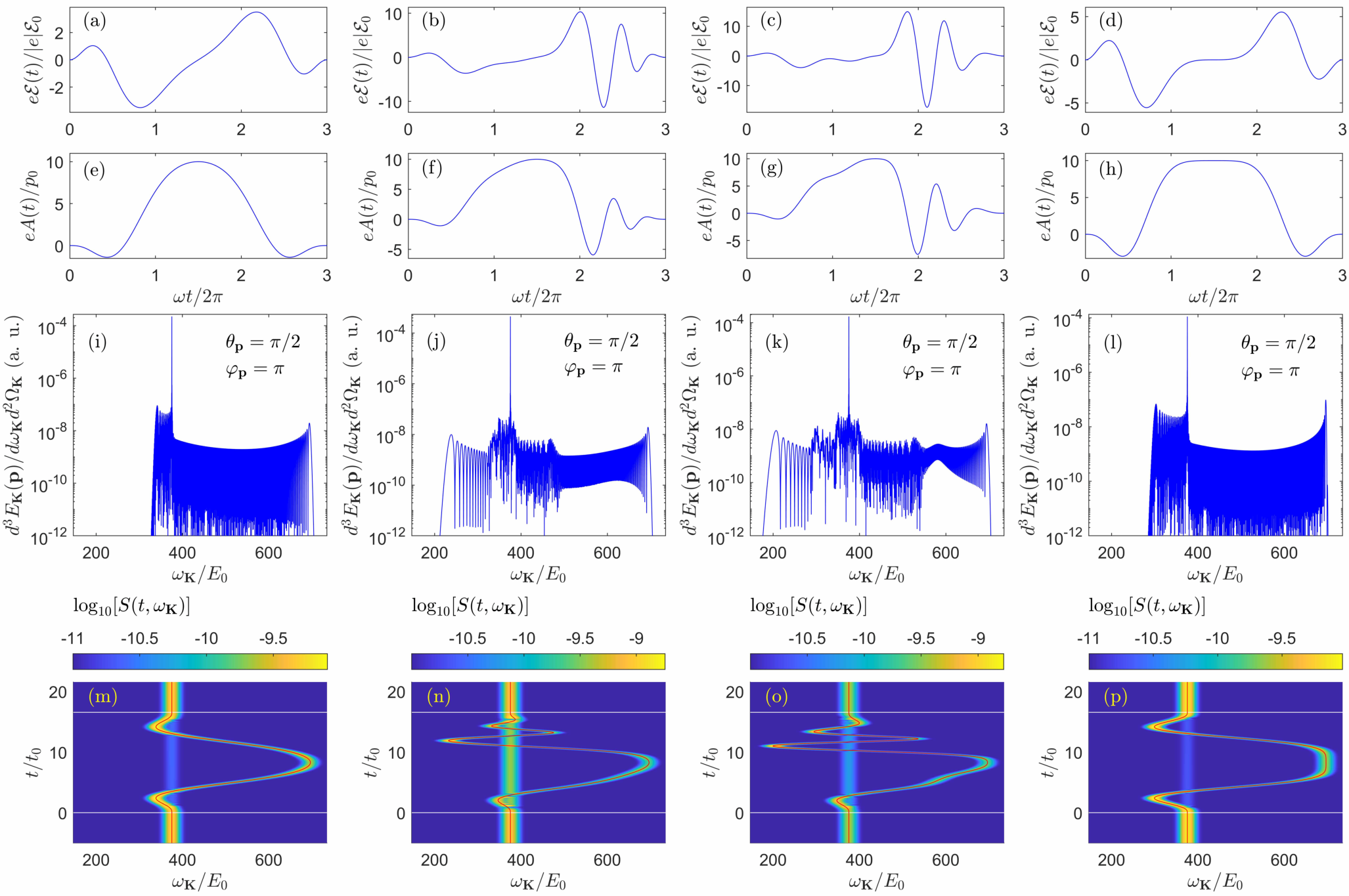}
\end{center}
\caption{Electron wave packet with the initial central energy $E_{\bm p}=10$~keV and a small momentum spread ($\Delta|{\bm p}|=2.74\times 10^{-5}p_0$) 
propagates along the $x$-axis and recombines with a hydrogen-like ion (${\cal Z}=4$)
to irradiate energy. The process is accompanied by a chirped laser pulse, that is linearly polarized in the $x$-direction and propagates along the $z$-axis.
Electric field (top row) and vector potential (second row) representing model chirped laser pulses, considered in Sec.~\ref{control}.
The third row presents the corresponding energy distributions of LARR radiation [Eq.~\eqref{a58}], with their spectrograms [Eq.~\eqref{c6}]
shown in the bottom row. The results are for the vector potential described by the shape function $f_1(t)$ [Eq.~\eqref{d1}] with 
$N_{\rm c}=0$ (first column), $N_{\rm c}=1$ (second column), $N_{\rm c}=2$ (third column), and by the shape function $f_2(t)$
[Eq.~\eqref{d2}] with $N_{\rm c}=0$ (last column). The remaining pulse parameters are: $\omega=1.14E_0$, $eA_0=10p_0$, $N_{\rm osc}=3$,
$\eta_0=-\frac{1}{6\pi}$, and $\chi=0$. It is assumed that the LARR radiation is linearly polarized along the $x$-axis and emitted
in the $z$-direction.
The spectrograms have been calculated using the Gaussian window function [Eq.~\eqref{c4}], characterized
by the parameters: $\xi_T=0.1, \xi_W=0.03, \omega_1=-5E_0$, and $\omega_2=21.55E_0$. In the bottom row, the red line represents the analytic
expression for the temporal irradiated energy $\omega_{\bm K}(t)$ [Eq.~\eqref{n3}], whereas the horizontal lines mark the beginning and the 
end of the laser pulses.}
\label{fig4}
\end{figure*}

In panel (c) of Fig.~\ref{fig2}, we present the variations of the LARR energy distributions with respect to the incident electron polar 
angle $\theta_{\bm p}$ near the high-energy cutoff. The solid line represents the high-energy portion of the distribution presented in panel (a), 
i.e., for $\theta_{\bm p}=0.432\pi$. The dashed line is for $\theta_{\bm p}=0.48\pi$, whereas the dotted line corresponds to $\theta_{\bm p}=0.5\pi$.
Hence, we observe that the plateau extends toward lower energies with increasing the polar angle $\theta_{\bm p}$. This agrees with Eq.~\eqref{n3}.
Based on this formula, we see that for $\theta_{\bm p}=0.5\pi$, the recoil correction vanishes as the electron is incident perpendicularly 
to the pulse propagation direction. In addition, the recoil correction in Eq.~\eqref{n3} is positive when ${\bm n}\cdot {\bm p}>0$.
In this case, i.e., for $\theta_{\bm p}\in(0,0.5\pi)$, the recoil correction does increase the LARR energy $\omega_{\bm K}^{(0)}(t)$ cutoff. In contrast,
in the case when ${\bm n}\cdot {\bm p}<0$, meaning for $\theta_{\bm p}\in(0.5\pi,\pi)$, the irradiated energy cutoff will be smaller as compared to that of
$\omega_{\bm K}^{(0)}(t)$. One can also check that for the angles considered in Fig.~\ref{fig2}(c)
the recoil correction in Eq.~\eqref{n3} is monotonically decreasing function of $\theta_{\bm p}$. Finally, we observe that for the current
parameters the plateau energy range has varied by roughly $8E_0$, which is by almost 220~eV. This indicates that nondipole effects in LARR
can be quite pronounced and have to be carefully accounted for in light of potential applications.

As it has been indicated by the above analysis, the longitudinal momentum transfer from the laser field plays a crucial role in determining properties
of LARR radiation. This is confirmed by Fig.~\ref{fig3}, where we present the complete polar mappings of the LARR energy spectra near the
high-energy cutoff. The results are for the geometry and parameters specified in Fig.~\ref{fig2}. While panel (a) has been calculated 
based on the approach that includes all leading-order nondipole corrections, in panel (b) we have accounted only for the retardation whereas
in panel (c) only for the recoil correction. Note that all mappings consist of interference pattern, with small and large probability regions
laying interchangeably. However, while the pattern in panel (b) peaks at $\theta_{\bm p}=0.5\pi$ and is symmetric with respect to that angle, 
in other cases the pattern is asymmetric, with the maximum being shifted toward smaller values of the electron polar angle $\theta_{\bm p}$. In order to explain this behavior,
we go back to Eqs.~\eqref{n3} and~\eqref{n4}. According to our analysis, for as long as the electron recoil off the laser pulse is not accounted for,
the temporal LARR energy is governed by Eq.~\eqref{n4}. Therefore, it depends on the electron polar angle through the term ${\bm p}\cdot{\bm A}\sim\sin\theta_{\bm p}$, 
which carries over to the distribution presented in panel (b). With an inclusion of the recoil correction [panels (a) and (c)], the electron acquires
longitudinal momentum kick from the laser pulse. It appears, therefore, as if it was incident with a smaller polar angle $\theta_{\bm p}$, that 
is measured with respect to the pulse propagation direction. As a result, we observe
asymmetric distributions of LARR radiation, with the aforementioned shift toward smaller values of $\theta_{\bm p}$.

In this section, we have investigated qualitatively new features of the LARR energy and angular distributions, which arise when the laser pulse
exerts radiation pressure on the recombining electron. Specifically, we have shown that the radiation pressure results in extension of the LARR high-energy 
plateau spectrum. Complementary to that, we shall study now whether the laser pulse can be tailored such that the LARR energy spectrum is enhanced.

\subsection{Control of LARR radiation by chirped laser pulses}
\label{control}

We consider the radiative recombination in the presence of a chirped linearly polarized laser pulse, that propagates in the $z$-direction
and is described by the vector potential,
\begin{equation}
{\bm A}(t)=A_0f_j(t){\bm e}_x,\quad j=1,2,
\label{d0}
\end{equation}
with the pulse shape given by either
\begin{align}
f_1(t)=\left\{ \begin{array}{ll}\sin^2\left(\frac{\omega t}{2N_{\rm osc}}\right)\sin[\varphi(t)],& 
\mbox{if\, $0\leqslant t\leqslant T_p$},\\
0, & \mbox{otherwise},\end{array} \right.
\label{d1}
\end{align}
or
\begin{equation}
f_2(t)=[f_1(t)+1]^2-1.
\label{d2}
\end{equation}
In addition, we assume that both pulses have nonlinear temporal phase,
\begin{equation}
\varphi(t)=\omega t+\chi+\eta_0(\omega t)^2\left[\sin\left(\frac{\omega t}{2N_{\rm osc}}\right)\right]^{2N_{\rm c}}.
\label{d3}
\end{equation}
Here, $\chi$ is for the carrier-envelope phase (CEP), whereas parameters $\eta_0$ and $N_{\rm c}$ characterize the laser field temporal chirp.
Specifically, when $\eta_0\neq 0$, the field has a linear chirp for $N_{\rm c}=0$, a chirp proportional to its instantaneous amplitude 
for $N_{\rm c}=1$, and to its instantaneous intensity for $N_{\rm c}=2$. These three cases are considered for the shape function $f_1(t)$ [Eq.~\eqref{d1}], as represented
in Fig.~\ref{fig4} in the first three columns. The remaining column in Fig.~\ref{fig4} relates to the pulse shape $f_2(t)$ [Eq.~\eqref{d2}]
with a linear temporal chirp. For each pulse, we set the CEP equal to $\chi=-[\pi N_{\rm osc}+\eta_0 (\pi N_{\rm osc})^2+\pi/2]$ modulo $2\pi$. 
With such choice of $\chi$, the vector potential for each pulse (when multiplied by the electron charge) peaks at $t=T_p/2$, as demonstrated in the second row of Fig.~\ref{fig4}. Note that the corresponding electric field 
${\cal E}(t)=-\partial_t A(t)$ is plotted in the top row. In addition, we have chosen $\eta_0=-1/(2\pi N_{\rm osc})$ which results in a rather
wide maximum of the vector potential characterized by the shape function $f_1(t)$. Most importantly, for such values of $\chi$ and $\eta_0$
the shape function $f_2(t)$ has the first three derivatives vanishing at $t=T_p/2$. Thus, causing the vector potential flattening in its central
region [Fig.~\ref{fig4}(h)]. Note that for the data presented in Fig.~\ref{fig4}, we have taken three-cycle laser pulses ($N_{\rm osc}=3$) with
the frequency $\omega=1.14E_0$ ($\lambda=40$~nm), and the amplitude equal to $eA_0=10p_0$.

\begin{figure}[t]
\centering
  \includegraphics[width=0.9\linewidth]{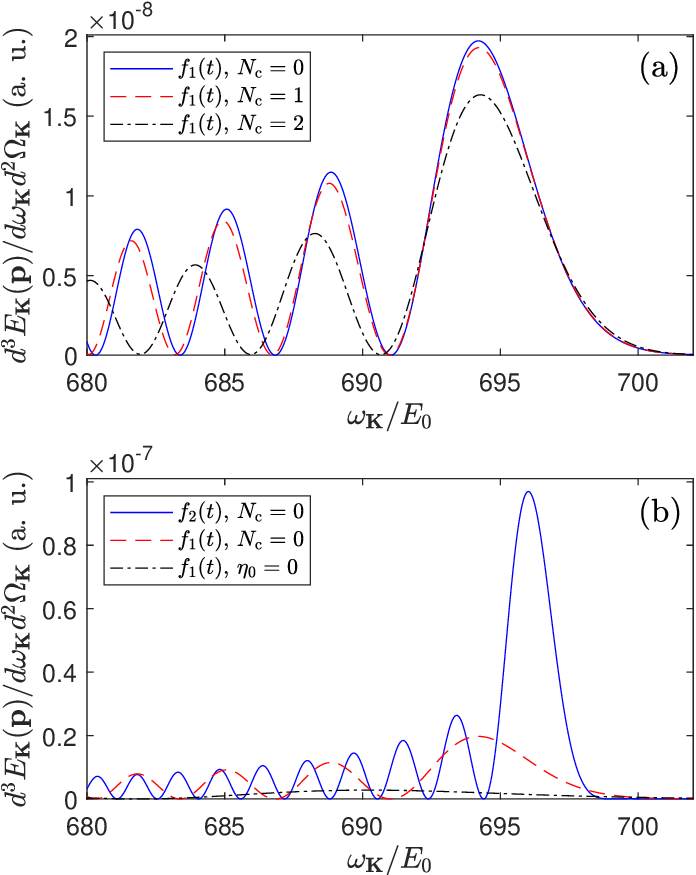}
\caption{Energy distributions of emitted radiation near the cutoff for various models of laser pulse assisting the radiative recombination.
The respective shape functions of those pulses are specified in the legend. The remaining parameters are the same as in Fig.~\ref{fig4}.}
\label{fig5}
\end{figure}

\begin{figure}[t]
\centering
  \includegraphics[width=0.95\linewidth]{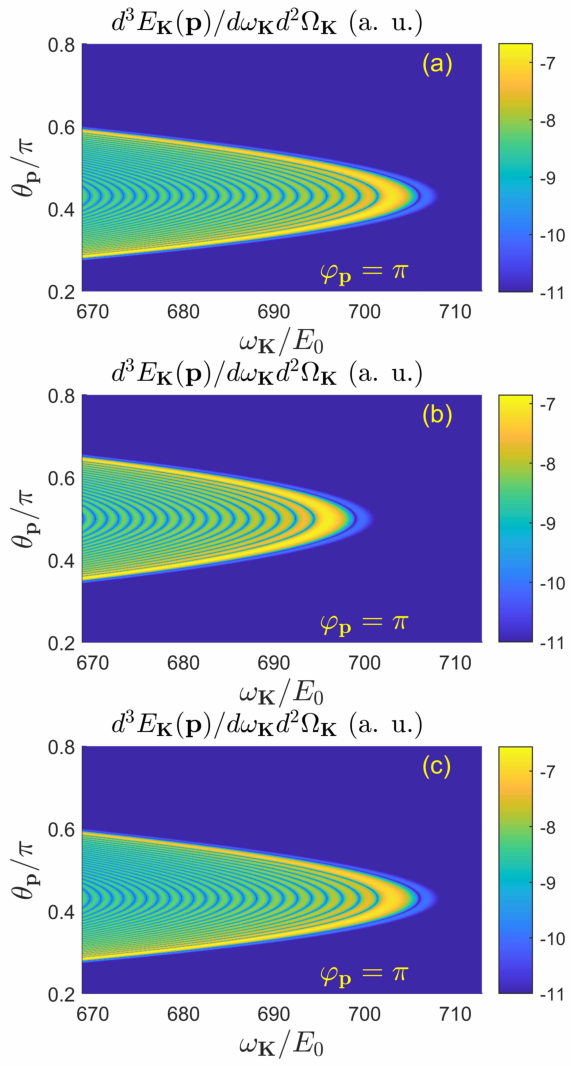}
\caption{Same as in Fig.~\ref{fig3} but for the parameters given in Fig.~\ref{fig4}. More specifically, the presented mappings were calculated
for the vector potential described by the shape function $f_2(t)$ [Eq.~\eqref{d2}] with $N_{\rm c}=0$, and for
$\omega=1.14E_0$, $eA_0=10p_0$, $N_{\rm osc}=3$, $\eta_0=-\frac{1}{6\pi}$, $\chi=0$. 
}
\label{fig6}
\end{figure}

In Fig.~\ref{fig4}, we present the energy spectra of LARR radiation emitted with polarization ${\bm \varepsilon}_{\bm K}={\bm e}_x$ in the $z$-direction when the aforementioned
model pulses assist the radiative recombination. As before, this happens when an electron described by a wave packet with the central initial
energy $E_{\bm p}=10$~keV and a small momentum spread, $\Delta |{\bm p}|=2.74\times 10^{-5}p_0$, is moving at angles $\theta_{\bm p}=\pi/2$ and $\varphi_{\bm p}=\pi$ to collide
with a hydrogen-like ion of ${\cal Z}=4$. We note that, except of overall differences, the end-portions of the spectra in the first three columns
of Fig.~\ref{fig4} seem to have similar magnitudes. To make it more clear, in Fig.~\ref{fig5}(a) we present their enlarged portions around the cutoff. 
As one can see, for the shape function $f_1(t)$ [Eq.~\eqref{d1}] the emission of LARR radiation becomes less efficient with increasing the parameter 
$N_{\rm c}$, even though for $N_{\rm c}=0$ (solid line) and $N_{\rm c}=1$ (dashed line) the magnitudes of the spectra are nearly the same. 
Going back to Fig.~\ref{fig4}, it is clear that the spectrum shown in the last column is significantly enhanced near the cutoff as compared to the other spectra.
This is confirmed in Fig.~\ref{fig5}(b), where we compare the end portion of the distributions for
pulses with linear chirps described by either the shape function $f_2(t)$ (solid line) or $f_1(t)$ (dashed line), and the pulse with the shape 
function $f_1(t)$ but no chirp ($\eta_0=0$). To make such comparison meaningful, for the latter we have assumed that the CEP equals
$\chi=-(-1)^{N_{\rm osc}}\frac{\pi}{2}$. In this case, the pulse maximum coincides with the maxima of chirped pulses. This in turn implicates, based on
Eq.~\eqref{n3}, that the cutoff for high energy radiation is the same in all presented cases. Still, the LARR spectrum for the shape function 
$f_2(t)$ is pronouncedly enhanced at the highest energies. To explain this, we note that the highest energy radiation is emitted when
the recombining electron picks up the maximum energy from the laser pulse, in accordance with Eq.~\eqref{n3}. This favors the pulse characterized
by the vector potential shown in Fig.~\ref{fig4}(h). In this case, emission of the highest-energy radiation is a compound effect, as the vector potential 
acquires the maximum value over the entire period. This makes the generation of the cutoff radiation most efficient.
Hence, we conclude that a proper shaping of the laser pulse assisting the radiative recombination permits to maximize 
the probability distributions of emitted radiation in the highest energy region. 

While the results presented in Figs.~\ref{fig4} and~\ref{fig5} have been calculated using the framework derived in Sec.~\ref{theory}, for the considered
geometry the nondipole effects were not very pronounced. For this reason, in Fig.~\ref{fig6} we show the energy distributions of LARR radiation versus the polar angle
of the incident electron $\theta_{\bm p}$ close to the LARR cutoff in the case when the laser pulse with a flat-top vector potential
[Fig.~\ref{fig4}(h)] is applied. The parameters are the same as in Fig.~\ref{fig4}, except that the electron polar angle $\theta_{\bm p}$ varies, 
as shown in the figure. Similar to Fig.~\ref{fig3}, we show the results accounting for all leading nondipole corrections [panel (a)] as well as
their individual contributions, coming from the retardation [panel (b)] and the recoil [panel (c)]. As one can observe, the main features of those
distributions are preserved. It is worth noting, however, that despite same laser peak ponderomotive energies used in relation to Figs.~\ref{fig3} 
and~\ref{fig6} the respective distributions are by an order of magnitude more pronounced for the latter. As we have already argued, this is
associated with the certain shape of the laser pulse, that facilitates a very efficient generation of the highest-energy radiation. Interestingly,
while the distributions presented in Fig.~\ref{fig6} show the same pattern as those in Fig.~\ref{fig3}, the regions of small and large probability
form a much denser structure in Fig.~\ref{fig6}. This means that for a flat-top pulse the energy distributions oscillate with energy more rapidly, which is also
visible in Fig.~\ref{fig5}(b). The interpretation of this effect can be based on the saddle-point approach presented in Sec.~\ref{nondipole}.
As it follows from Fig.~\ref{fig4}(p), in the high-energy region there are two saddle points contributing to the LARR probability amplitude.
However, as the pulse has a flat top, those saddle points are quite well-separated in time. Hence, by superposing their corresponding contributions,
one obtains very fast oscillations of probability distributions in the energy domain. Once again, we note that the laser pulse that assists 
the radiative recombination provides important means to control the process.

\section{Conclusions}
\label{conclusions}

We have demonstrated a comprehensive theoretical description of electron-atom radiative recombination in the presence of a laser pulse 
that accounts for nondipole corrections. The latter were derived systematically by the relativistic reduction, enabling to follow the origin 
of various nondipole contributions. As it followed from our numerical analysis, the most important of them originates from the electron
recoil off the laser pulse. We have recognized that, for the parameters considered in the paper, it results in extension of the LARR high-energy 
plateau and is a cause of an asymmetry of energy distributions of generated radiation with respect to the polar angle of recombining electron. 
The remaining nondipole corrections of the leading order in $1/c$ turned out to be less significant.

In the present paper, we have analyzed ways to control the LARR radiation using external laser pulses. Despite the aforementioned nondipole 
effects which arise entirely from the presence of the laser field, we have also investigated a possibility of manipulating the intensity 
of the cutoff portion of LARR by chirping the accompanying laser pulse. More specifically, we have been able to shape the laser pulse such that
the recombining electron acquires maximum energy from the field for an extended period of time. This has allowed us to maximize the intensity of the high-energy
LARR radiation, which can be of interest while synthesizing ultrashort laser pulses out of LARR.

\section*{Acknowledgements}

This work is supported by the National Science Centre (Poland) under Grant Nos.~2018/30/Q/ST2/00236 (D.K., M.M.M., K.K.) 
and~2018/31/B/ST2/01251 (J.Z.K.), and by the National Natural Science Foundation of China under Grant Nos.~12234002 and~92250303
(L.-Y.P.).

\appendix
\section{Boca-Florescu transformation}
\label{app}

Let us consider a regularized integral,
\begin{equation}
I_\varepsilon=\int_{-\infty}^\infty\dd t\,\ee^{-\varepsilon |t|}\ee^{\ii Q t}F(t),
\label{b1}
\end{equation}
where $\varepsilon\rightarrow 0^+$. Moreover, in relation to Eq.~\eqref{a33}, we assume that the function $F(t)$ has the following structure,
\begin{equation}
F(t)=\ee^{\ii H(t)}f(t),
\label{b2}
\end{equation}
and the following conditions are satisfied:
\begin{eqnarray}
H(t)=&0,\;\;&{\rm for}\;\; t<0,\nonumber\\
H(t)=&\,H(T_p),\;\;&{\rm for}\;\; t>T_p,\nonumber\\
f(t)=&f(0),\;\;&{\rm for}\;\; t<0\;\; {\rm and}\;\; t>T_p.
\label{b3}
\end{eqnarray}
In our case, $T_p$ corresponds to the laser field duration and $f(0)$ is the field-free portion of $f(t)$. Taking into account 
the above conditions, one can rewrite Eq.~\eqref{b1} as
\begin{equation}
I_\varepsilon=\frac{F(0)}{\ii(Q-\ii\varepsilon)}+\int_0^\infty\dd t\,\ee^{\ii(Q+\ii\varepsilon)t}F(t),
\label{b4}
\end{equation}
where the integral in Eq.~\eqref{b1} was divided into two intervals, $(-\infty,0]$ and $[0,\infty)$, and the former one was calculated
by parts. Now, in order to calculate the remaining integral, we use the identity,
\begin{equation}
\int_0^\infty\dd t\,\frac{\dd}{\dd t}\left[\ee^{\ii(Q+\ii\varepsilon)t}F(t)\right]=-F(0),
\label{b5}
\end{equation}
which leads to
\begin{align}
\int_0^\infty\dd t\,&\ee^{\ii(Q+\ii\varepsilon)t}F(t)=-\frac{F(0)}{\ii(Q+\ii\varepsilon)}\nonumber\\
&-\frac{1}{\ii(Q+\ii\varepsilon)}\int_0^{T_p}\dd t\,\ee^{\ii(Q+\ii\varepsilon)t}\dot{F}(t).
\label{b6}
\end{align}
Here, the range of integration in the last integral has been limited to $[0,T_p]$ because $\dot{F}(t)=0$ for $t>T_p$. Plugging this result in Eq.~\eqref{b4},
one obtains therefore
\begin{align}
I_\varepsilon=\frac{F(0)}{\ii(Q-\ii\varepsilon)}&-\frac{F(0)}{\ii(Q+\ii\varepsilon)}\nonumber\\
&-\frac{1}{\ii(Q+\ii\varepsilon)}\int_0^{T_p}\dd t\,\ee^{\ii(Q+\ii\varepsilon)t}\dot{F}(t).
\label{b7}
\end{align}
Taking the limit $\varepsilon\rightarrow 0^+$ in Eq.~\eqref{b7}, we obtain
\begin{align}
I=\lim_{\varepsilon\rightarrow 0^+}I_\varepsilon&=\pi\delta(Q)[F(0)+F(T_p)]\nonumber\\
&+\ii{\cal P}\left(\frac{1}{Q}\right)\int_0^{T_p}\dd t\,\ee^{\ii Qt}\dot{F}(t),
\label{b8}
\end{align}
where we have applied the Sokhotski-Plemelj theorem,
\begin{equation}
\lim_{\varepsilon\rightarrow 0^+}\frac{1}{Q\pm\ii\varepsilon}={\cal P}\left(\frac{1}{Q}\right)\pm\ii\pi\delta(Q),
\label{b88}
\end{equation}
with ${\cal P}(\cdot)$ denoting the principle value. Realizing that
\begin{equation}
F(0)+F(T_p)=2f(0)\ee^{\ii H(T_p)/2}\cos\left(H(T_p)/2\right),
\label{b9}
\end{equation}
and writing the explicit form of $\dot{F}(t)$, which follows from Eq.~\eqref{b2},
Eq.~\eqref{b8} becomes
\begin{align}
I&=2\pi\delta(Q)f(0)\ee^{\ii H(T_p)/2}\cos\left(H(T_p)/2\right)\nonumber\\
&+\ii{\cal P}\left(\frac{1}{Q}\right)\int_0^{T_p}\dd t\,\ee^{\ii Qt+\ii H(t)}\Bigl[\dot{f}(t)+\ii \dot{H}(t)f(t)\Bigr].
\label{b10}
\end{align}
As argued in Sec.~\ref{amplitude}, this allows us to identify the field-free and the field-modified contributions to the probability amplitude of LARR.

\section{Classical considerations}
\label{classical}

The aim of this Appendix is to derive the formula for the kinetic energy of an electron moving in an electromagnetic field, within the framework of classical electrodynamics. We consider an electron of momentum ${\bm p}$
moving in an electromagnetic field that propagates in the direction ${\bm n}$. The latter represents a finite laser pulse that is described by the electric and magnetic field components, 
${\bm{\mathcal{E}}}({\bm r},t)$ and ${\bm{\mathcal{B}}}({\bm r},t)$, respectively. Note that for a transverse field it holds that ${\bm n}\cdot {\bm{\mathcal{E}}}({\bm r},t)=0$
and ${\bm n}\cdot {\bm{\mathcal{B}}}({\bm r},t)=0$. Hence, we proceed with the Newton-Lorentz equation for the electron motion,
\begin{equation}
\dot{\bm \pi}=e{\bm{\mathcal{E}}}({\bm r},t)+\frac{e}{m_{\rm e}}{\bm \pi}\times {\bm{\mathcal{B}}}({\bm r},t),
\label{ddd1}
\end{equation}
where ${\bm \pi}$ is the electron kinetic momentum such that
\begin{equation}
m_{\rm e}\dot{\bm r}={\bm \pi}
\label{ddd111}
\end{equation}
and
\begin{equation}
{\bm{\mathcal{B}}}({\bm r},t)=\frac{1}{c}{\bm n}\times{\bm{\mathcal{E}}}({\bm r},t).
\label{ddd2}
\end{equation}
Plugging the above relation into Eq.~\eqref{ddd1} and using the triple product formula, we obtain
\begin{equation}
\dot{\bm \pi}=e{\bm{\mathcal{E}}}({\bm r},t)+\frac{e}{m_{\rm e}c}{\bm n}\left[{\bm \pi}\cdot {\bm{\mathcal{E}}}({\bm r},t)\right]
-\frac{e}{m_{\rm e}c}{\bm{\mathcal{E}}}({\bm r},t)({\bm n}\cdot{\bm \pi}).
\label{ddd3}
\end{equation}
Next, we expand Eq.~\eqref{ddd3} in powers of $1/c$, leaving only the leading order terms,
\begin{equation}
\dot{\bm \pi}=e{\bm{\mathcal{E}}}(t)-\frac{{\bm n}\cdot{\bm r}}{c}e\dot{\bm{\mathcal{E}}}(t)
+\frac{e}{m_{\rm e}c}{\bm n}\left[{\bm \pi}\cdot {\bm{\mathcal{E}}}(t)\right]
-\frac{e}{m_{\rm e}c}{\bm{\mathcal{E}}}(t)({\bm n}\cdot{\bm \pi}).
\label{ddd4}
\end{equation}
We look for solutions of this equation in the form
\begin{equation}
{\bm \pi}={\bm p}-e{\bm A}(t)+\delta{\bm \pi}(t),
\label{ddd5}
\end{equation}
where ${\bm p}$ is the dynamical electron momentum that is defined asymptotically, whereas ${\bm A}(t)$ denotes the vector potential describing 
the electromagnetic field in the dipole approximation such that ${\bm{\mathcal{E}}}(t)=-\dot{\bm A}(t)$. Moreover, $\delta{\bm \pi}(t)$ represents 
the leading order correction (of the order of $1/c$) to the electron kinetic momentum in the field, ${\bm p}-e{\bm A}(t)$. Inserting
Eq.~\eqref{ddd5} into Eq.~\eqref{ddd4}, we obtain up to the first order in $1/c$ that
\begin{eqnarray}
\delta\dot{\bm \pi}(t)=&-&\frac{{\bm n}\cdot{\bm r}}{c}e\dot{\bm{\mathcal{E}}}(t)+\frac{e}{m_{\rm e}c}{\bm n}\cdot\left[({\bm p}
-e{\bm A}(t))\cdot{\bm{\mathcal{E}}}(t)\right]\nonumber\\
&-&\frac{e}{m_{\rm e}c}{\bm{\mathcal{E}}}(t)\cdot({\bm n}\cdot{\bm p}).
\label{ddd6}
\end{eqnarray}
Here, we have used the fact that ${\bm n}\cdot{\bm A}(t)=0$ for a transverse field. The solution to this equation has the form,
\begin{equation}
\delta{\bm \pi}(t)=-\frac{{\bm n}\cdot{\bm r}}{c}e{\bm{\mathcal{E}}}(t)-\frac{e}{m_{\rm e}c}{\bm n}\cdot\left[{\bm p}\cdot{\bm A}(t)\right]
+\frac{e^2{\bm A}^2(t)}{2m_{\rm e}c}{\bm n}.
\label{ddd7}
\end{equation}
Together with Eq.~\eqref{ddd5}, it allows us to derive the classical formula for the kinetic energy of an electron in an electromagnetic field, 
which accounts for the leading nondipole corrections. Namely,
\begin{eqnarray}
\frac{{\bm \pi}^2}{2m_{\rm e}}&=&\frac{\left[{\bm p}-e{\bm A}(t)\right]^2}{2m_{\rm e}}-\frac{{\bm n}\cdot{\bm p}}{m_{\rm e}^2c}
\Big[e\bm{A}(t)\cdot\bm{p}-\frac{1}{2}e^2\bm{A}^{2}(t)\Big]\nonumber\\
&+&\frac{{\bm n}\cdot{\bm r}}{m_{\rm e}c}\left[({\bm p}-e{\bm A}(t))\cdot e{\bm{\mathcal{E}}}(t)\right]+{\cal O}\Big(\frac{1}{c^2}\Big).
\label{ddd8}
\end{eqnarray}
While the second term in Eq.~\eqref{ddd8} corresponds to the electron recoil, the third term is the reminiscence of retardation. 
We note that for the electron in an electromagnetic field, ${\bm n}\cdot{\bm r}\sim {\bm n}\cdot [{\bm p}-e{\bm A}(t)]/m_{\rm e}={\bm n}\cdot{\bm p}/m_{\rm e}$,
which follows from Eqs.~\eqref{ddd111} and~\eqref{ddd5} for a transverse field.
Therefore, this correction is negligible for as long as the electron is impinging nearly perpendicularly to the field propagation direction,
when ${\bm n}\cdot{\bm p}\approx 0$. This is in agreement with our results and their interpretation, as presented in Sec.~\ref{nondipole}.

\section{Spectrogram}
\label{spect}

In order to calculate the spectrogram $S(t,\omega_{\bm K})$ of a signal ${\cal A}(\omega)$, the latter is first truncated to a finite interval 
$\omega\in[\omega_1, \omega_2]$. This is to avoid the Gibbs effect, as we have discussed in Ref.~\cite{Deeksha}. Then, we define the truncated signal 
${\cal A}_T(\omega)$ as
\begin{equation}
{\cal A}_T(\omega)={\cal A}(\omega)f_T\left(\frac{\omega-\omega_1}{\omega_2-\omega_1},\xi_T (\omega_2-\omega_1)\right),
\label{c1}
\end{equation}
where $\xi_T$ is a small parameter and 
\begin{equation}
f_T(x,\Delta x)= \left\{ \begin{array}{ll}
0, & \textrm{for $x\leqslant 0$},\\
\sin^2\left(\frac{\pi x}{2\Delta x}\right), & \textrm{for $0<x<\Delta x$},\\
1, & \textrm{for $ \Delta x\leqslant x\leqslant 1-\Delta x $},\\
\sin^2\left(\frac{\pi (1-x)}{2\Delta x}\right), & \textrm{for $1-\Delta x<x<1$},\\
0, & \textrm{for $x\geqslant 1$}.
\end{array} \right.
\label{c2}
\end{equation}
The short-time Fourier transform of the truncated signal is defined as
\begin{equation}
{\cal A}_{ST}(t,\omega_{\bm K})=\int_{\omega_1}^{\omega_2}\dd\omega{\cal A}_T(\omega)W(\omega-\omega_{\bm K},\xi_W(\omega_2-\omega_1))\ee^{-\ii\omega t},
\label{c3}
\end{equation}
where $\xi_W$ is a parameter determining the width of the window function $W(x,\Delta x)$. Moreover, for calculations presented in this paper
we shall use the Gaussian window,
\begin{equation}
W(x,\Delta x)=\frac{\ee^{-(x/\Delta x)^2}}{\sqrt{\pi}\Delta x},
\label{c4}
\end{equation}
satisfying the condition,
\begin{equation}
\int_{-\infty}^\infty\dd xW(x,\Delta x)=1.
\label{c5}
\end{equation}
Finally, we define the spectrogram as~\cite{Deeksha}
\begin{equation}
S(t,\omega_{\bm K})=|{\cal A}_{ST}(t,\omega_{\bm K})|^2,
\label{c6}
\end{equation}
that provides information about the frequency spectrum of the signal while it varies in time.




\end{document}